\documentclass[twocolumn,showpacs,amsmath,amssymb]{revtex4}

\usepackage{graphicx}
\usepackage{dcolumn}
\usepackage{bm}

\begin{document}


\title{A Quantum Computer Architecture using Nonlocal Interactions\\}

\author{Gavin K. Brennen, Daegene Song, and Carl J. Williams}
\affiliation{National Institute of Standards and Technology, Gaithersburg, Maryland 20899-8423\\
}


\begin{abstract}
Several authors have described the basic requirements essential to
build a scalable quantum computer.  Because many physical
implementation schemes for quantum computing rely on nearest
neighbor interactions, there is a hidden quantum communication
overhead to connect distant nodes of the computer.  In this paper
we propose a physical solution to this problem which, together
with the key building blocks, provides a pathway to a scalable
quantum architecture using nonlocal interactions.  Our solution
involves the concept of a quantum bus that acts as a refreshable
entanglement resource to connect distant memory nodes providing an
architectural concept for quantum computers analogous to the von
Neumann architecture for classical computers.
\end{abstract}


\maketitle
Most modern computers share the same basic architecture
first proposed by von Neumann in 1945.  Von Neumann organized a
computer into four basic components:  memory, an input/output
system, an arithmetic logic unit, and a control unit.  The four
units were interconnected by a bus that provided for the flow of
classical bits or information between the various components
\cite{1}.  Basic elements sufficient to build a scalable quantum
computer have been described by DiVincenzo \cite{2}  and  Preskill
\cite{3}.  The five DiVincenzo criteria \cite{2} for building a
quantum computer are: a scalable physical system with well
characterized qubits, the ability to initialize the state of the
qubits to a simple fiducial state, long relevant decoherence
times, a universal set of quantum gates, and a qubit specific
measurement capability.  In addition, Preskill lists other
elements necessary for fault tolerant computation in order to
maintain a reasonable accuracy threshold.  Two of these are
maximal parallelism and gates that can act on any pair of qubits.

Although Preskill \cite{3} communicates the need to interact
arbitrary pairs of qubits, he provides no solution for this in a
typical quantum computer restricted to nearest neighbor
interactions.  DiVincenzo \cite{2} mentions two additional
criteria essential for quantum communications namely:  the ability
to interconvert stationary and flying qubits, and the ability to
faithfully transmit flying qubits between specified locations.
Clearly, if such capabilities were engineered into the
architecture, the above requirements of qubit interconnectivity
and parallelism could simultaneously be satisfied.  In this paper
we show an alternative approach based on the concept of a quantum
bus that consists of refreshable qubits that act as a resource for
entanglement.  This concept bears similarity to the classical bus,
key to the von Neumann architecture.

For concreteness we consider a lattice model of a quantum computer
(e.g. a neutral atom optical lattice, quantum dot arrays, or
$^{31}$P embedded Si \cite{4}) where qubits are fixed in position
and interactions are with nearest neighbors (Fig. 1). One obvious
way to connect distant qubits is to swap the states through
intermediary qubits until the states are adjacent to each other,
perform the requisite operations, and then swap back.  This
procedure scales linearly with distance between the pair and the
resultant fidelity due to one and two qubit errors associated with
swapping falls off exponentially with distance.  One can make this
fault tolerant by swapping through an ancillae at each step using
fault tolerant CNOT gates, however, this requires a physical
architecture that can accommodate sufficient numbers of ancillae
between any two memory qubits.  The consequence is that swapping
can introduce large overhead in terms of computational steps and
ancillae when one includes error correction on the swapping gates
themselves and on the quantum memory of the computer during the
operations.

In contrast, our approach is to divide the physical qubits of the
computer into static domains storing quantum memory and a dynamic
bus of qubits connecting the domains (Fig. 1).  Nearest neighbor
pairs within the bus can be entangled by spatially selective
system interaction.  By performing measurement at the joints
between the pairs the entanglement can be swapped \cite{5} to
provide entanglement between the ends of the bus.  Any nonlocal
two qubit controlled unitary gate can then be implemented using
one maximally entangled bus pair neighboring the distant memory
nodes using only nearest neighbor operations and classical
communication \cite{6,7}.  This approach using entanglement
swapping has the advantage that the ``quantum bus" need not meet
the same requirements as fault tolerant computation but must only
reach the minimal threshold required for entanglement purification
\cite{8}.  Note that this model for a quantum bus using nearest
neighbor interactions differs from a common quantum bus shared
between all memory nodes as in the ion trap quantum computing
proposals \cite{9}.

The efficacy of this protocol depends on the ordering of numerous
time scales including gate times for one- and two-qubit
operations, measurement times, and decoherence times.  In our
approach, error rates are divided into static decoherence errors
for errors that occur when a qubit is {\it not being manipulated}
and dynamic decoherence that results from manipulating the qubits.
The described architecture is appropriate to the situation where
the static decoherence time is much longer than the other time
scales in the problem so that the limitation on fidelity of the
computation is due almost completely to dynamical, one- and
two-qubit, errors.  Our proposal also requires Bell state
measurements on the joints between nearest neighbor entangled
pairs, consequently the measurement errors, $\epsilon_{meas}$,
must be comparable to dynamical errors.  The requirements on the
error rates must therefore satisfy $\epsilon_{meas}\sim\epsilon_{2
bit,1 bit}\gg\epsilon_{static}$.

\begin{figure}
\includegraphics[scale=0.35]{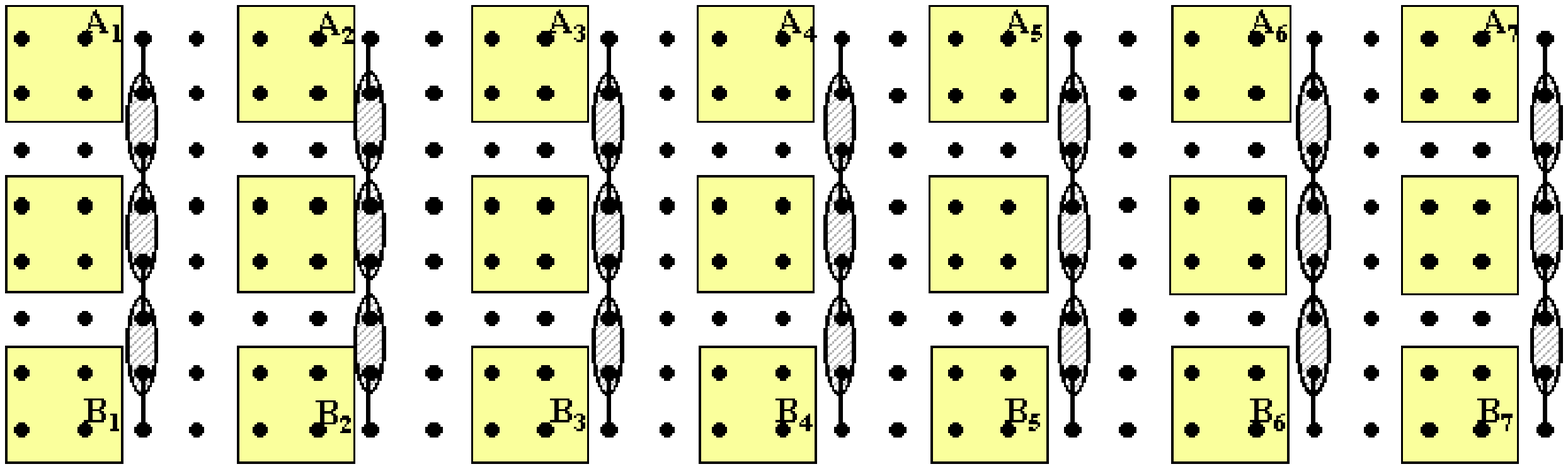}
\caption{\label{fig:1}A lattice model of a quantum computer. The
qubits in the boxes correspond to static physical qubits storing
quantum memory, here shown comprising a 7 qubit quantum error
correction code \cite{10}.  Refreshable dynamic qubits in the
channel are used as a bus to carry information stored in the
static qubits.  Pairwise entanglement is generated along the bus
indicated by lines connecting physical qubits and Bell
measurements are made at the joints (ellipses) to perform
entanglement swapping.  By creating parallel entanglement
resources, nonlocal operations can be implemented transversally
between code blocks.  Increasing the number of qubits in a box can
accommodate concatenated encoding. }
\end{figure}

We first describe the procedure for implementing resource swapping
under ideal operations and then discuss the effect of noise on the
protocol.  Resource swapping with nearest neighbor interactions
involves entanglement swapping through $l-1$ qubits beginning with
$l$ unentangled bus qubits and ending with a distant entangled
pair $\rho_{1,l}$.  The protocol can be realized by performing two
sets of two-qubit and one-qubit gates in parallel followed by
measurement and a single qubit completion gate on one end as
diagrammed in Fig. 2. While this procedure takes $l/2-1$ Bell
measurements, unlike swap operations discussed above, all the
measurements can be performed simultaneously instead of in $O(l)$
steps \cite{11}.  The completion gate $\sigma_{i,j}^{1}$, where
$\sigma_{0,0} \equiv {\bf 1}, \sigma_{0,1}\equiv \sigma_x ,
\sigma_{1,0}\equiv \sigma_z, \sigma_{1,1}\equiv -i \sigma_{y}$,
transforms the four possible maximally entangled Bell states
$|\Psi^{i,j}\rangle=\sigma_{i,j}^{1\dagger}(|00\rangle+|11\rangle)/\sqrt{2}$
resulting from the measurement into the fiducial state
$|\Psi^{0,0}\rangle$.  Because the Pauli operators anticommute,
the completion gate depends only on the parity of measurement
results, $m_{j}\in\{0,1\}$, over even and odd ordered qubits:
$\sigma_{M} = \prod_{j=1}^{l/2-1} \sigma_{m_{2j},m_{2j+1}} \equiv
\sigma_{\oplus_j m_{2j},\oplus_j m_{2j+1}}$.  There are two
important features of the completion gate.  First, because it only
acts on the first qubit, it commutes with all other operations in
the entanglement swapping meaning all intermediate measurements
can be made simultaneously.  Second, the completion gate depends
only on the bitwise sums of the even and odd qubits between the
distant pair.  As such if the entanglement bus is set up in an
alternating order of physically distinct species, then it is only
necessary to collect two classical bits of information: a parity
measurement of the even and odd (e,o) indexed species.  If a
detector can discriminate parity for each species, e.g. in the
case of atomic systems by counting parity of scattered photons
from transitions on $|1\rangle_{e,o}$ to excited states
$|f\rangle_{e,o}$ induced by two resonant fields, then one need
only have global addressability of the two species and local
addressability at the boundaries.  For instance, a lattice
architecture could be built with some addressable impurity or
boundary near each memory qubit location, relaxing the constraint
of addressability along the intervening channels.  An example of a
system that could exploit this parallelism is proposed in
\cite{12} where counter-propagating beams of cross polarized light
produce an alternating array of potential wells trapping atoms of
two species with polarization along $\sigma_{+}$ or $\sigma_{-}$.
Rotating the relative angle of polarization allows selective
pairwise interactions with left or right neighbors of a particular
species.

Maximally entangled EPR pairs can be used as a resource for
perfect nonlocal gates, obviating the need for swapping memory
qubits.  In a real experimental setup there will be noise in this
process due to imperfect  control over one- and two-qubit unitary
operations as well as measurement errors.  The resulting distant
entangled pair after noisy operations will be in a mixed state
whose character depends on the noise and the measurement results
of the intervening states.  We focus on physical systems wherein
single qubit unitary operations can be implemented with
near-perfect fidelity.  This is the case, for instance, in many
quantum optical systems such as ion traps, cavity QED, and optical
lattices \cite{4}.  A counterexample is in liquid state NMR
\cite{13} wherein the many qubit coupling gates are always on and
careful pulse engineering is needed to implement one-qubit gates
selectively.  In principle, all one-qubit errors could be
incorporated into the two qubit error map except for the final
single bit completion gate (Fig. 2).  We consider two types of two
qubit errors.  One is depolarizing error described by the map:
\begin{equation}
S_{DEP}(\rho)=p\,U\rho U^{\dagger}+(1-p)\,Tr_{i,j}[\rho ]\otimes {\bf 1}_{i,j}/4,\;
\end{equation}
where $U$ is the desired unitary to be performed on qubits ($i,j$)
(in our case $U$=CPHASE) and $p$ is the probability that the gate
was successful.  This error can be interpreted as a process where
with a probability $1-p$ one of the sixteen possible combinations
of tensor products of two single qubit unitaries chosen uniformly
and randomly from the set
$\{\sigma_{\alpha}\otimes\sigma_{\beta}\}$
($\alpha,\beta=0,x,y,z$) acts on the interacting qubits during the
expected gate.  This kind of error can occur when there is
uncertainty of the control fields modeled as additional single bit
rotations sampled from an isotropic distribution acting on the two
qubits.

\begin{figure}
\begin{center}
\includegraphics[scale=0.35]{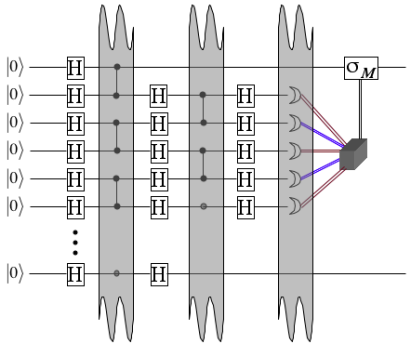}
\caption{\label{fig:2}Quantum circuit to implement entanglement
swapping in six time steps,  independent of length.  The bus
qubits are initialized to $|0\rangle$ and subsequent one-qubit,
$H=\sigma_{z}e^{i\frac{\pi}{4}\sigma_{y}}$, and two-qubit gates,
CPHASE$=e^{i\pi|11\rangle\langle11|}$, are applied in parallel.
The shaded gray time slices correspond to periods where two-qubit
noise or measurement error may occur.  The lines connecting to the
classical processor represent classical information from
measurement results on qubits of alternating ordered species that
need not be individually addressed.}
\end{center}
\end{figure}

A second error model is the controlled phase error:
\begin{equation}
S_{CPE}(\rho)=\int \,d\phi\,g(\phi)\,U(\phi) \rho
U(\phi)^{\dagger}\;
\end{equation}
where $U(\phi)=e^{i (\phi+\pi)|11\rangle\langle11|}$, is the
CPHASE gate with an additional unknown phase sampled from the
probability distribution $g(\phi)$.  In the case that $g(\phi)$ is
symmetric about zero, the map is simply
\begin{equation}
S_{CPE}(\rho)=p\,U(0)\rho U(0)^{\dagger}+(1-p)\,\rho. \;
\end{equation}
This map corresponds to a physical situation where some
experimental uncertainty in the field strength, timing, or
strength of the interaction, imparts an additional unwanted phase
during the gate.  An example where this can occur is in the
proposals for controlled phase gates using dipole-dipole
interactions between trapped alkalis \cite{12,14}.  In these
proposals, fluctuations in the trapping potential or dipole
inducing laser amplitude or detuning results in a nonseparable
phase accumulation.

To account for measurement error we associate the experimental
measurement outcomes 0 and 1 with two dimensional projectors,
$P_{0}=\eta|0\rangle\langle0|+(1-\eta)|1\rangle\langle1|$, and
$P_{1}=\mbox{\boldmath$1$}-P_{0}$. This model includes less than
perfect detector efficiency $\eta$ since there is a probability
$(1-\eta)$ that a detector reading of 0 actually results from the
qubit being in state $|1\rangle$, and conversely for 1.  In many
systems the efficiency can be improved at the cost of lengthening
measurement time.  For example, the internal state of atoms or
ions can be detected by optically pumping population to
``stretched states" of maximal spin angular momentum projection
and tuning to a resonant transition with the excited state.
Presence and absence of scattered photons corresponds to a zero or
a one and detector inefficiency due to dark counts can be
suppressed by scattering more photons.  In this way efficiencies
of $0.9999$ can be obtained \cite{15}.

The effect of entanglement swapping through one pair of qubits
under the depolarizing map produces a Bell diagonal state. The
recursion relation can be solved to show that the state after
swapping through $n=l/2-1$ pairs is:
\begin{equation}
\begin{array}{lll}
\rho_{1,l}&=&p^{l-1} \sigma^{1}_{M}
(a_{+ n}|\Psi^{0,0}\rangle\langle\Psi^{0,0}|+b_{n}(|\Psi^{0,1}\rangle\langle\Psi^{0,1}|\\
& &+|\Psi^{1,0}\rangle\langle\Psi^{1,0}|)+a_{- n}|\Psi^{1,1}\rangle\langle\Psi^{1,1}|)\sigma^{1 \dagger}_{M}\\
& &+(1-p^{l-1})\,{\bf 1}/4,\;
\label{rhofinal}
\end{array}
\end{equation}
where $a_{\pm n}=1/4 (1\pm2(2\eta-1)^n+(2\eta-1)^{2n})$,
$b_{n}=(1-a_{+ n}-a_{- n})/2$, and $\sigma^{1}_{M}$ is the
completion gate on qubit 1.  The fidelity for a length $l$ pair is
defined as the overlap of the resulting state with the maximally
entangled state
$F_{l}=\langle\Psi^{0,0}|\sigma^{1}_{M}\rho_{1,l}\sigma^{1
\dagger}_{M}|\Psi^{0,0}\rangle$.  For depolarizing error the
fidelity is
\begin{equation}
F_{l}(p,\eta)=\frac{1}{4} (1+p^{l-1}(2(2\eta-1)^{(l-1)/2}+(2\eta-1)^{l-1}).\;
\end{equation}

The CPE map during entanglement swapping creates a mixed state
that is a convex sum of Bell diagonal and logical basis diagonal
states.  The recursive map for this model has a complicated form
as a function of number of swaps but it is straightforward to show
that the fidelity $F_{l}$ is the same as that for the depolarizing
error.  Indeed, upon randomizing the state after all measurements
with the twirl \cite{8} operator
$\mathcal{T}(\rho)=1/4\sum_{\alpha=0}^{3}{\sigma_
{\alpha}\otimes\sigma_{\alpha}\rho\sigma^{\dagger}_{\alpha}\otimes\sigma^{\dagger}_{\alpha}}$,
the state is equal to Eq. \ref{rhofinal}.  We can generalize this
error model to include the effect of leakage due to coherent
evolution that takes population out of the logical basis.  As a
simplification we assume this process occurs only for population
in the $|11\rangle$ state such that during the CPHASE gate
population coherently evolves into states $|k\rangle$ outside the
logical basis: $|11\rangle\rightarrow
-e^{i\phi-\gamma/2}|11\rangle+\sum_{k}{a_{k}|k\rangle}$. Tracing
over the other states, the effective evolution can be related to
the CPE model with nonunitary evolution by $U(\phi+i\gamma)$ where
$U(\phi)$ is as above and $\gamma$ is an effective decay.  The CPE
with leakage does not display some of the nice symmetry properties
of the other two error models and the fidelity as a function of
number of swaps does not have a general closed form solution.
Under the assumption of independent Gaussian noise on the
additional phase $\phi$ in $U(\phi)$ and in the limit that the
probability of error over the number of swaps is small
($l\gamma,l(1-p)\ll1$), the fidelity is approximately
\begin{equation}
\begin{array}{lll}
F_{l}(\gamma,p,\eta)&\approx&\frac{1}{16}(4 p^{l-1}e^{-l\gamma}((2\eta-1)^{l-1}\\&
&+2(2\eta-1)^{(l-1)/2})+3+e^{-2l\gamma}).\;
\end{array}
\end{equation}

It is evident that for the error models considered here the
fidelity falls off exponentially with distance, however, as long
as the measurement error is not too large, the fidelity ratio of
swapping information versus resource swapping is exponentially
small.  This is evident because it requires at least $l$ swaps to
connect two qubits a distance $l$ apart and each swap requires at
least 2 maximally entangling gates meaning $F^{SWAP}_{l}<p^{2l}$.
The time to implement parallel entanglement swapping is $T_{ent
swap}=4\tau_{1 bit}+2\tau_{2 bit}+\tau_{meas}$ \emph{independent
of length} and can be much faster than the minimal swapping time
$T_{swap}=2l\tau_{2 bit} $ provided $\tau_{meas}$ is not too
large.

Ultimately, in order to perform a high fidelity nonlocal gate, the
long distant mixed entangled pairs will have to be purified. There
are several protocols for entanglement purification. Efficient
protocols which use two way classical communication work by
performing nearest neighbor operations at each end of two mixed
state pairs and, based on measurement results on each of two
particles in a target pair, the round succeeds and the control
pair's fidelity improves, or the round fails and measurement
results on both pairs are disregarded.  Provided the initial pairs
have fidelity above a certain threshold ($F_{min}>1/2$ for perfect
operations), the map will converge to $F_{max}=1$ after a finite
number of rounds.  D\"{u}r \emph{et al.} {\cite{16} demonstrate
that by using quantum repeaters, one can achieve high fidelities
with noisy operations while sacrificing a number of qubit
resources that scales polynomially with the length of the channel
and a subsequent time cost.

If a wide entanglement bus with many parallel channels is
available in a quantum computer architecture, the quantum
repeaters nesting algorithm of entanglement swapping and
purification actions may be preferable to a single purification
stage as used in the Deutsch protocol \cite{17}.  It will depend
on the time scales for single qubit memory decoherence times
whether the additional time cost of the repeaters is overall
advantageous for robust quantum information processing.  As an
example, given a measurement detector efficiency $\eta=0.99$, a
two qubit gate success probability $p=0.995$, and no decay, a
length $l=25$ entangled pair can be made with fidelity $F=0.74$.
After six successful rounds of entanglement purification under the
Deutsch protocol the resultant single pair will have fidelity
$F=0.985$.  The initial entangled pairs can be made in parallel
and pairs can be nested inside each channel so that the bus
between adjacent memory qubits need not be too wide.

Once the state $\rho_{1,l}$ has been purified to an acceptable
fidelity, then the nonlocal gate can be implemented between two
memory qubits $A$ and $B$ using nearest neighbor gates between $A$
and $1$ and $B$ and $l$ and measurement on the qubits $1$ and $l$
\cite{7}.  For a given resource in a Bell diagonal state,
$\rho_{1,l}=a|\Psi^{0,0}\rangle\langle\Psi^{0,0}|+b|\Psi^{1,0}\rangle\langle\Psi^{1,0}|
+c|\Psi^{0,1}\rangle\langle\Psi^{0,1}|+d|\Psi^{1,1}\rangle\langle\Psi^{1,1}|$,
where $a>b,c,d$, the fidelity of the gate is determined by the
ability of this resource to map a product state of $A,B$ to a
maximally entangled state and is given by
\begin{equation}
F_{gate}(p,\eta)=p^{2}(a\eta^{2}+(b+c)\eta(1-\eta)+d(1-\eta)^{2})+(1-p^{2})/4.\;
\end{equation}

The quantum bus architecture described in this paper appears to be
appropriate to neutral atoms and the nuclear spin version of
${^{31}P}$ embedded in Si, for example.  This is because the
dominant decoherence in these situations is believed to result
from imperfect one- and two- qubit operations, and not due to
static memory decoherence; in contrast to some schemes using
SQUIDS and quantum dots \cite{4} where errors appear to be as
likely at times between gates as during gates.  The projected
solution is attractive because information does not need to be
moved thereby reducing memory decoherence and the overall clock
time for the nonlocal gate operations.  Although we describe our
model in terms of a 2D lattice of qubits it could be readily
extended to a multiplexed set of ion traps.  Also, 3D lattices or
alternative 2D lattices such as hexagonal close packing may be
advantageous, especially with regard to resource scheduling.  For
instance, some quantum algorithms can be parallelized to exploit
the commutivity of certain operations \cite{18} if pairs of memory
nodes can be simultaneously connected.  It is not clear what the
optimal scheme is to create the necessary entanglement resources
simultaneously to perform such nonlocal operations since the
resource of bus qubits is limited.  Nevertheless, it is apparent
that a 3D lattice will have a significant advantage over the 2D
with internode distances that scale like $l^{1/3}$ vs. $l^{1/2}$,
and diverse pathways for resource swapping.

This work was supported in part by DARPA under the QuIST program
and by ARDA/NSA.  We would like to acknowledge Mark Heiligman for
first suggesting the idea of quantum architectures and for useful
discussions.


\begin{references}
\bibitem{11} We assume $l$ even, however, for $l$ odd one can
follow the same procedure with $l-1$ qubits and a single swap
operation at one end.
\end{references}

\end{document}